\documentclass{PoS}

\title{Composite Taus and Higgs Decays}

\ShortTitle{Composite Taus and Higgs Decays}

\author{Adri\'an Carmona$^\ast$\\
Institute for Theoretical Physics, \\
ETH Zurich, 8093 Zurich, Switzerland\\
        E-mail: \email{carmona@itp.phys.ethz.ch}}
\author{Florian Goertz\\
Institute for Theoretical Physics, \\
ETH Zurich, 8093 Zurich, Switzerland\\
        E-mail: \email{fgoertz@itp.phys.ethz.ch}}

\author{\vspace{-9.5mm}\phantom{\speaker{Adri\'an Carmona and Florian Goertz}}}

\abstract{ 
In this talk, we describe the effects of extended fermion sectors, respecting custodial symmetry, on Higgs production and decay. The resulting protection for the $Z \to b_L \bar b_L$ and $Z \to \tau_R \bar \tau_R$ decays allows for potentially interesting signals in Higgs physics, while maintaining the good agreement of the Standard Model with precision tests. The setups can be motivated as the low energy effective theories of the composite Higgs models MCHM$_5$ and MCHM$_{10}$, where the spectra can be identified with the light custodians present in these theories. We will show that these describe the
relevant physics in the fermion sectors in a simplified and transparent way. In contrast to previous studies of composite models, the impact of a realistic lepton sector on the Higgs decays is taken into account.
}

\FullConference{The European Physical Society Conference on High Energy Physics -EPS-HEP2013\\
		18-24 July 2013\\
		Stockholm, Sweden}

\begin{document}

\vspace*{-12.4mm}
\section{Introduction}
\vspace{-2.4mm}

The properties of the Higgs-like particle discovered at the LHC are so far found to be in good agreement with the Standard Model (SM) predictions. 
On the other hand, there is also still room for sizable deviations, in particular in the decay into bottom quarks, $\tau$ leptons, and photons.
Vector-like fermion scenarios offer viable and simple extensions of the SM with an interesting phenomenology for these Higgs decay modes. Especially 
new leptons respecting custodial symmetry allow to substantially modify the latter two decays, without generating a conflict with precision tests \cite{Carmona:2013cq}.

In the following, we will examine on the one hand the impact of simple and transparent (low-energy) vector-like fermion scenarios, featuring a custodial symmetry 
to protect the $Z b_L \bar b_L$ and $Z \tau_R\bar{\tau}_R$  couplings, on Higgs physics at the LHC. Beyond that, we will however also 
address the question where these fermions come from and study strongly coupled UV completions of the scenarios, delivered
by different minimal composite Higgs models (MCHM)~\cite{Agashe:2004rs}. 
%MCHM$_5$ and MCHM$_{10}$.  Here, the subscripts stand for the representations of the fermions under the $SO(5)$ symmetry 

Due to the fact that in such models there exist fermionic resonances associated to the heavy top quark (and $\tau$ lepton), required by custodial symmetry, 
that are typically much lighter than the actual scale of these models, $m_{\rm cust} \ll f$, their low energy theories match just the aforementioned vector-like 
fermion setups. Although, due to the smaller masses, there is \emph{a priori} no reason to expect such light custodians also in the lepton sector, note that 
explaining the observed pattern of lepton masses and mixings with the help of a discrete $A_4$ symmetry requires the $\tau$ to be more composite than naively 
expected and thus triggers the appearance of light $\tau$ custodians \cite{delAguila:2010vg}. At low energies $E\ll f$ one would thus see the SM spectrum 
plus resonances coming with the top quark and the $\tau$ lepton, which just corresponds to what we will be studying. The additional (sub-leading) impact 
of the  UV completions will be detailed at the end, concentrating on the most important effect of the non-linearity of the Higgs-sector. Due to the 
compositeness of the $\tau$ one expects non-negligible effects in the lepton sector of Higgs physics, which had been neglected in the past literature. 
\vspace{-1mm}

\section{Setup}
\vspace{-2.4mm}

The emergence  of light $\tau$ custodians within the framework of the full MCHM$_5$ has been motivated in \cite{delAguila:2010vg} % from a UV perspective
% for a complete composite Higgs model,
and they persist also when the $\tau_R$ partners transform in a \textbf{10} of $SO(5)$~\cite{next}. 
However, as mentioned before, the setup for our analysis of Higgs production and decay will just rely on the corresponding low energy theories, including the light custodians. For arbitrary fermionic representations, this low energy spectrum will not necessarily be a good description of the full theory of gauge Higgs unification (GHU), as the \emph{a priori} suppressed contributions of heavier modes can be lifted by larger Yukawa couplings. Nevertheless, for the models we will consider here, the full theory is indeed effectively described by the effects of the light custodians~\cite{Carmona:2013cq}. In particular, we~will study the cases where the composite partners of the $\tau_R$ transform in fundamental or adjoint representations of $SO(5)$, while all other composite fermions span fundamental representations, which we will denote by MCHM$_5$ and MCHM$_{5+10}$, even if we only study the low energy setups.
\vspace{-1mm}

\subsection{MCHM$_5$}
\label{sec:MCHM5}
\vspace{-1mm}

The light $\tau$ custodians present in this model in addition to the SM fermions are contained in the following degenerate lepton doublets \cite{delAguila:2010es}
\begin{eqnarray}
L_{1L,R}^{(0)}=\left(\begin{array}{c}N_{1L,R}^{(0)}\\ E_{1L,R}^{(0)}\end{array}\right) \sim \mathbf{2}_{-\frac{1}{2}},\quad 
L_{2L,R}^{(0)}=\left(\begin{array}{c} E_{2L,R}^{(0)}
\\ Y_{2L,R}^{(0)} \end{array}\right)\sim \mathbf{2}_{-\frac{3}{2}}\,,
\label{eq:2Doub}
\end{eqnarray}
where we are explicitly giving the quantum numbers under the electroweak gauge group $SU(2)_L\times U(1)_Y$.  The superscript $(0)$ indicates the current basis. The model is designed such that the custodial symmetry can protect the $Z \tau_R \bar \tau_R$ coupling via a $P_{LR}$ symmetry. In addition, we assume a similar embedding of the quark sector \cite{Carmona:2013cq}, with the replacement $N \to \Lambda,\,E \to T$ and $Y \to B$ and hypercharges 7/6 and 1/6 respectively,  now featuring a protection for the $Z b_L \bar b_L$ coupling. Neglecting the first two generations, which are assumed to have negligible couplings to the new resonances, the mass and Yukawa couplings for the lepton sector are given by 
\begin{eqnarray}
{\cal L}_{L} =
- y_l\, \bar{l}_L^{(0)}\varphi \tau^{(0)}_R
-y_l^\prime
\Big[\bar{L}^{(0)}_{1L} \varphi + \bar{L}^{(0)}_{2L}
  \tilde{\varphi} \Big] \tau^{(0)}_R 
- M_l \Big[ \bar{L}^{(0)}_{1L} L^{(0)}_{1R}+ \bar{L}^{(0)}_{2L}
  L^{(0)}_{2R} \Big]+\mathrm{h.c.},
  \label{eq:Ll}
\end{eqnarray}
where $\varphi$ is the Higgs doublet. 
A similar expression holds for the quark sector. We have
 neglected the couplings of the right handed neutrino (bottom
quark), as they are SM like since they do not have any new resonance to couple to.

After diagonalizing the different mass matrices,  the Higgs couplings with leptons become \cite{Carmona:2013cq}
\begin{eqnarray}
\label{eq:gh5}
g_{h5}^E=U_L^{5\dagger}(s_L)\ g_{h5}^{E(0)}\, U_R^5(s_R)=\frac 1 v
\left(\begin{array}{ccc}
c_R^2 m_\tau & 0 & s_R c_R m_\tau \\
0 & 0 & 0 \\
s_R c_R M_{E_2}
& 0 & 
s_R^2 M_{E_2}
\end{array}\right)\,,
\end{eqnarray}
where $m_{\tau}, M_{E_1}$, and $M_{E_2}$ are the physical  $Q=-1$ lepton masses and $s_{L,R}=\sin(\theta_{L,R})$, $c_{L,R}=\cos(\theta_{L,R})$ are the sine and the cosine of the mixing angles. It turns out that $s_R$, measuring the mixing of the  $\tau_R$ with the resonances, together with one of the  $M_{E_i}$ can be taken as the only free parameters of the model. Analogous couplings arise for the $T$ sector \cite{Carmona:2013cq}. Note that the other new resonances present in the lepton and quark sectors do not couple to the Higgs boson at all. 
\vspace{-1mm}

\subsection{MCHM$_{5+10}$}
\vspace{-1mm}

In this case, as the composite partners of the $\tau_R$ transform in an adjoint representation of $SO(5)$, which decomposes under $SU(2)_L\times SU(2)_R$ as $\mathbf{10}=(\mathbf{2},\mathbf{2})+(\mathbf{3},\mathbf{1})+(\mathbf{1},\mathbf{3})$, the lepton spectrum will have  on top of the bidoublet spanned by  (\ref{eq:2Doub}) one additional $SU(2)_L$ triplet and two singlets 
\begin{eqnarray}
	L_{3L,R}^{(0)}&=&\left(\begin{array}{c} N_{3L,R}^{(0)}\\E_{3L,R}^{(0)}
\\ Y_{3L,R}^{(0)} \end{array}\right)\sim \mathbf{3}_{-1}, \quad N_{2L,R}^{(0)}\sim \mathbf{1}_0,\quad Y_{1L,R}^{(0)}\sim \mathbf{1}_{-2}.
\end{eqnarray}
The relevant part of the Yukawa and mass Lagrangian for this case can be found in \cite{Carmona:2013cq}.  The rotations to the mass basis will be analogous to the previous case but now featuring larger
matrices.  However, in this case the $Q=-2$ vector-like leptons $Y_i$ will have non vanishing Yukawa couplings to the Higgs boson, leading to a very distinct phenomenology compared to the MCHM$_5$.
\vspace{-1mm}

\section{Higgs Production and Decay}
\label{sec:Higgs}
\vspace{-1.8mm}

%\begin{figure}[!t]
%	\begin{center}
%	\mbox{\includegraphics[height=5.5cm]{triangles.pdf}}
%	\vspace{-1cm}
%\parbox{15.5cm}{\caption{\label{fig:triangles}First row: Leading-order contribution to Higgs-boson production via gluon-gluon fusion and contribution 
%from heavy quark resonances to the same process; Second row: Leading contributions to the Higgs decay into two photons, given by a top-quark loop and 
%a $W^\pm$-boson loop, as well as contributions from heavy fermion resonances to the same process.}}
%\end{center}
%\vspace{0.3cm}
%\end{figure}
The presence of the new resonances has interesting implications on the production and decay of the Higgs boson, which will be worked out in this section. 
We will first concentrate on the MCHM$_5$ and later discuss the generalization to the MCHM$_{5+10}$.

The most important production mechanism for the Higgs boson at hadron colliders is gluon-gluon fusion, which in the SM receives its main contribution from a 
top-quark triangle loop. Here, the new quarks (the top custodians) give rise to additional loop diagrams, with contributions proportional to the corresponding Higgs coupling over the new quark masses
\begin{equation}\label{eq:nu5}
    \nu_T^5 \equiv v \;\sum_{n = 2}^3 \; \frac{{\rm Re}\left[ (g_{h5}^T)_{nn}\right]}{m_{T_{n-1}}} A_{q}^h(\tau_{T_{n-1}})\approx(s_R^t)^2 \,,
\end{equation}
where $\tau_i=4 m_i^2/m_h^2$.
In the second step we have already inserted the explicit result for our model, following from (\ref{eq:gh5}) and used the fact that the loop function $A_{q}^h$ approaches an asymptotic value of 1 for virtual particles much heavier than the Higgs, which is already a good approximation for the SM top quark. 
For details, see \cite{Carmona:2013cq}.
Moreover, there are changes in the couplings of the Higgs to SM-like quarks,
which for the top quark amount to a reduction of
\begin{equation}
\label{eq:kappaf}
\kappa_t^5\equiv v {\rm Re}\left[(g_{h5}^T)_{11}\right]/m_t=(c_R^t)^2\, ,
\end{equation}
and vanish for bottom quarks, $\kappa_b^5=1$.

%, with a large coupling to the Higgs, see the leftmost diagram in Figure \ref{fig:triangles}. In extensions of the SM this process can receive corrections from 
%new colored particles that propagate in the loop (see the second diagram in the figure) as well as from modified couplings 
%of the SM quarks in the loop to the Higgs boson. Both effects are present in the models we consider. 

The deviations in gluon fusion can now be parametrized  by a rescaling factor $\kappa_g^5$ as
\begin{equation}
\label{eq:gg}
\sigma(gg\to h)_{{\rm MCHM}_5}=|\kappa_g^5|^2\, \sigma(gg\to h)_{\rm SM}\,,
\end{equation}
\begin{equation} \label{eq:kappag} 
  \kappa_g^5 = \frac{{\displaystyle
      \sum}_{f = t, b} \, \kappa_f^5 \hspace{0.25mm} A_{q}^h
    (\tau_f)\, + \nu_T^5 }{ {\displaystyle \sum}_{f = t,
      b} \; A_{q}^h (\tau_f)} \approx \frac{(c_R^t)^2 \hspace{0.25mm}+ A_{q}^h(\tau_b)\, 
     + (s_R^t)^2}{ 1+A_{q}^h (\tau_b)}\,.
\end{equation}
Employing $(c_R^t)^2 + (s_R^t)^2 = 1$, we see explicitly that the new loop contributions from the top custodians cancel with the 
corrections to the SM-like loops and thus gluon fusion remains to leading approximation SM-like, i.e. $\kappa_g^5\approx 1$. This cancellation of the fermionic parameters agrees with the effects found in composite 
models \cite{Falkowski:2007hz}. Finally, the change in the $h t \bar t$ vertex also modifies $tth$ production, leading to $\sigma(tth)_{{\rm MCHM}_5}
=(\kappa_t^5)^2\, \sigma(tth)_{\rm SM}$, whereas other relevant production mechanisms are unaffected by the new fermions,
$\sigma({\rm VBF})_{{\rm MCHM}_5}=\sigma({\rm VBF})_{\rm SM}$ and $\sigma(V h)_{{\rm MCHM}_5}=\sigma(V h)_{\rm SM}$.

We now turn to the Higgs decays and want to analyze in particular if a similar cancellation that happens in gluon fusion also 
appears in the loop induced decay into photons.
We define
\begin{equation}
\label{eq:Gam}
\Gamma(h\to ff)_{{\rm MCHM}_5}=|\kappa_f^5|^2\, \Gamma(h\to ff)_{\rm SM}\,,
\end{equation}
$f=\gamma,W,Z,b,\tau,g$.
Since we did not change the bosonic sector of the SM so far, we get
$ \kappa_W^5=\kappa_Z^5=1$.
Beyond that, the rescaling factors for the (tree-level) decays into two fermions as well as the decay to gluons, entering (\ref{eq:Gam}),  
have already been specified in (\ref{eq:kappaf}), with obvious replacements for $\kappa_\tau^5$, and (\ref{eq:kappag}).
Note that in our setup the couplings of the light fermions to the Higgs boson are unchanged.\footnote{In (anarchic) composite models, this is a good approximation due to the tiny mixing with the fermionic resonances.}
%\footnote{In extra dimensional setups or composite Higgs models, with anarchic flavor structure, this is motivated by the fact that the first two generations have negligible interactions with the Kaluza-Klein 
%excitations, or composite fermions.}
For the effective coupling to photons we obtain
\begin{eqnarray} \label{eq:kappagamma} 
\kappa_\gamma^5 =&
  \frac{{\displaystyle \sum}_{f = t, b} \; N_c \hspace{0.5mm} Q_f^2
    \hspace{0.5mm} \kappa_f^5 \hspace{0.25mm} A_{q}^h (\tau_f) + Q_\tau^2 
    \kappa_\tau^5 \hspace{0.25mm} A_{q}^h (\tau_\tau) +
    A_W^h (\tau_W) +  N_c \hspace{0.5mm} Q_t^2
    \hspace{0.5mm} \nu_T^5+ Q_\tau^2
    \hspace{0.5mm} \nu_E^5 }{{\displaystyle \sum}_{f = t,
      b} \; N_c \hspace{0.5mm}  Q_f^2 \hspace{0.5mm} A_{q}^h (\tau_f) 
    +Q_\tau^2 \hspace{0.5mm} A_{q}^h (\tau_\tau) +
    A_W^h (\tau_W)}\\ \nonumber
    =&
  \frac{N_c (Q_t^2 +Q_b^2
    \hspace{0.5mm} A_{q}^h (\tau_b) )+ Q_\tau^2 (
    (c_R^\tau)^2 \hspace{0.25mm} A_{q}^h (\tau_\tau) +  
     (s_R^\tau)^2)  +
    A_W^h (\tau_W) }{N_c (Q_t^2+Q_b^2 \hspace{0.5mm} A_{q}^h (\tau_b) )
   +  Q_\tau^2 \hspace{0.5mm} A_{q}^h (\tau_\tau)  +  A_W^h (\tau_W)}
 \approx
  \frac{-5 + 
     (s_R^\tau)^2}{-5}\,,
\end{eqnarray}
where $N_c =3$, $Q_t = 2/3$, $Q_b =
-1/3$, $Q_\tau = -1$.
The first, second, and third terms in
the numerator in the first row describe the effects of virtual SM-type quark, lepton, and $W^\pm$-boson
exchange, respectively. The fourth and fifth term, on the other hand, correspond to the contributions of the custodians, which now also include leptonic resonances
- the lepton contribution $\nu_E^5$ is defined in analogy to (\ref{eq:nu5}).
Note that the amplitude proportional to $A_{W}^h (\tau_W)\approx -6.25$ dominates in the SM and interferes destructively 
with the fermion contributions $A_{q}^h (\tau_f)$. 
In the second row we have already employed that $(c_R^t)^2 + (s_R^t)^2 = 1$. One can see an interesting thing happening.
Due to the fact that the $\tau$ is much light than the Higgs (leading to $A_{q}^h (\tau_\tau)\ll 1$) a similar cancellation as for quarks is {\it not} happening in the lepton sector,
which includes the new case of a light fermion with nevertheless naturally a significant composite component. 
This leads to a reduction of the Higgs coupling to two photons, which can become quite significant and has been neglected before~\cite{Carmona:2013cq}. 

In the case of the MCHM$_{5+10}$,  besides a larger parameter space which calls for numerical methods of diagonalization and a random scan to study the phenomenological predictions of the model, we have the distinctive feature of new $Q=-2$ vector-like excitations coupling to the Higgs boson and therefore entering the $h\to \gamma \gamma$ loop process \cite{Carmona:2013cq}. On the other hand, as the gauge and quark sectors remain the same, we still have $\kappa_{W}^{5+10}$$=\kappa_Z^{5+10}$$=\kappa_{b}^{5+10} \approx \kappa_{g}^{5+10}\approx 1$ and $\kappa_{t}^{5+10}=\kappa_{t}^{5}$. 

\begin{figure}[!t]
\begin{center} 
\includegraphics[width=0.3\textwidth]{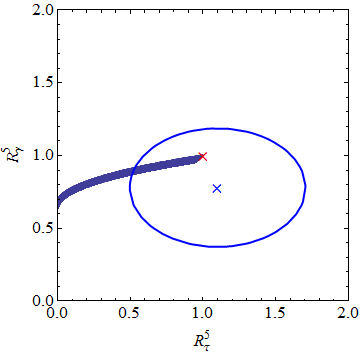}
\includegraphics[width=0.3\textwidth]{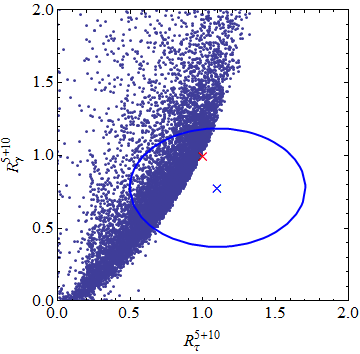}
\includegraphics[width=0.3\textwidth]{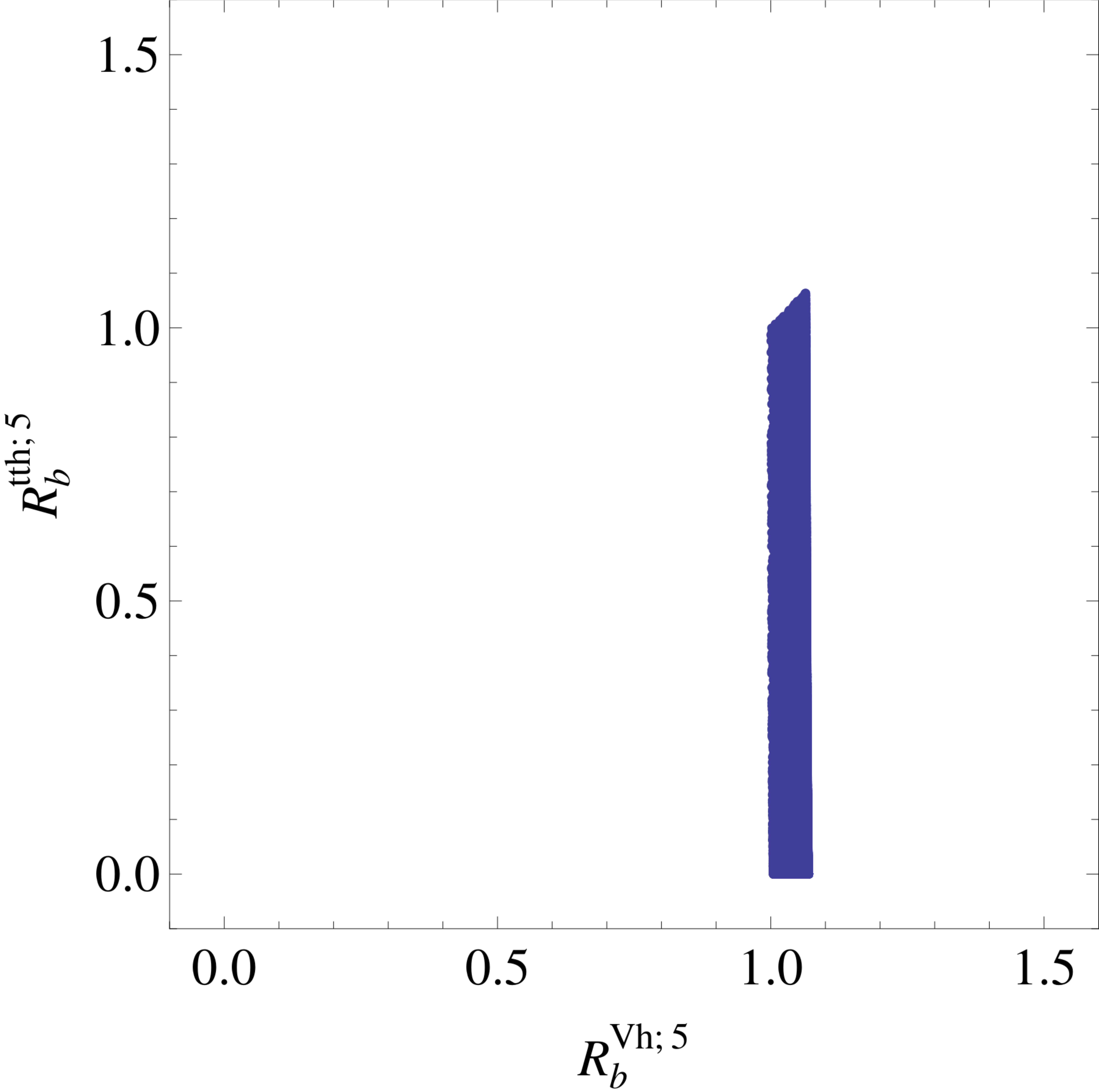}
\caption{\label{fig:gamgamtautau} Left and Middle: Production cross section times decay for $pp\to h\to \gamma\gamma$ versus $pp\to h\to \tau\tau$, relative to the SM, for both models considered. The ellipse sketches the CMS 1$\,\sigma$ bound. Right: $pp\to h\to bb$ via $tth$ production over the SM result versus the same ratio mediated by $Vh$. See text and \cite{Carmona:2013cq} for details.}
\end{center}
\vspace{-0.5cm}
\end{figure}

To  compare both models and  have an idea of the possible size of the deviations with respect to the SM, we show in the left and central panels of Figure~\ref{fig:gamgamtautau}  the ratio  of the production cross section times branching fraction for $pp\to h\to \gamma\gamma$ relative to the SM versus the same ratio for $pp\to h\to \tau\tau$ in both models (see \cite{Carmona:2013cq} for more details). While in the MCHM$_5$ we scan $0\leq s_R \leq 1$, in the MCHM$_{5+10}$ we have assumed random vector-like masses of $\sim (0.2-1)$\,TeV as well as order one Yukawa couplings with the strong sector, motivated from the UV completion. We again refer the reader to \cite{Carmona:2013cq} for more details. Comparing the plots,  we can see that the strong correlations present in the MCHM$_5$ are washed out in the MCHM$_{5+10}$, making on the one-hand side the model less predictive but on the other allowing in principle for either an excess or a suppression in $pp\to h\to \gamma\gamma$. However, in the GHU model completing the MCHM$_{5+10}$  there are additional correlations present between the parameters. If we impose them to mimic its UV completion,  it turns out that all the points corresponding to $R_\tau^{5+10}>1$ or  $R_\gamma^{5+10}>1$ in the middle panel of the previous figure disappear. We have also plotted, in the right panel of Figure~\ref{fig:gamgamtautau}, the Higgs production cross section in the $tth$ channel times branching fraction $h\to bb$ relative to the SM versus the equivalent ratio, assuming that the Higgs has been produced via associated vector-boson production. Once the experimental situation in these channels improves they will become a
superb tool for measuring directly a possible reduction in the $tth$ coupling.

\begin{figure}[!t]
\begin{center} 
\includegraphics[width=0.3\textwidth]{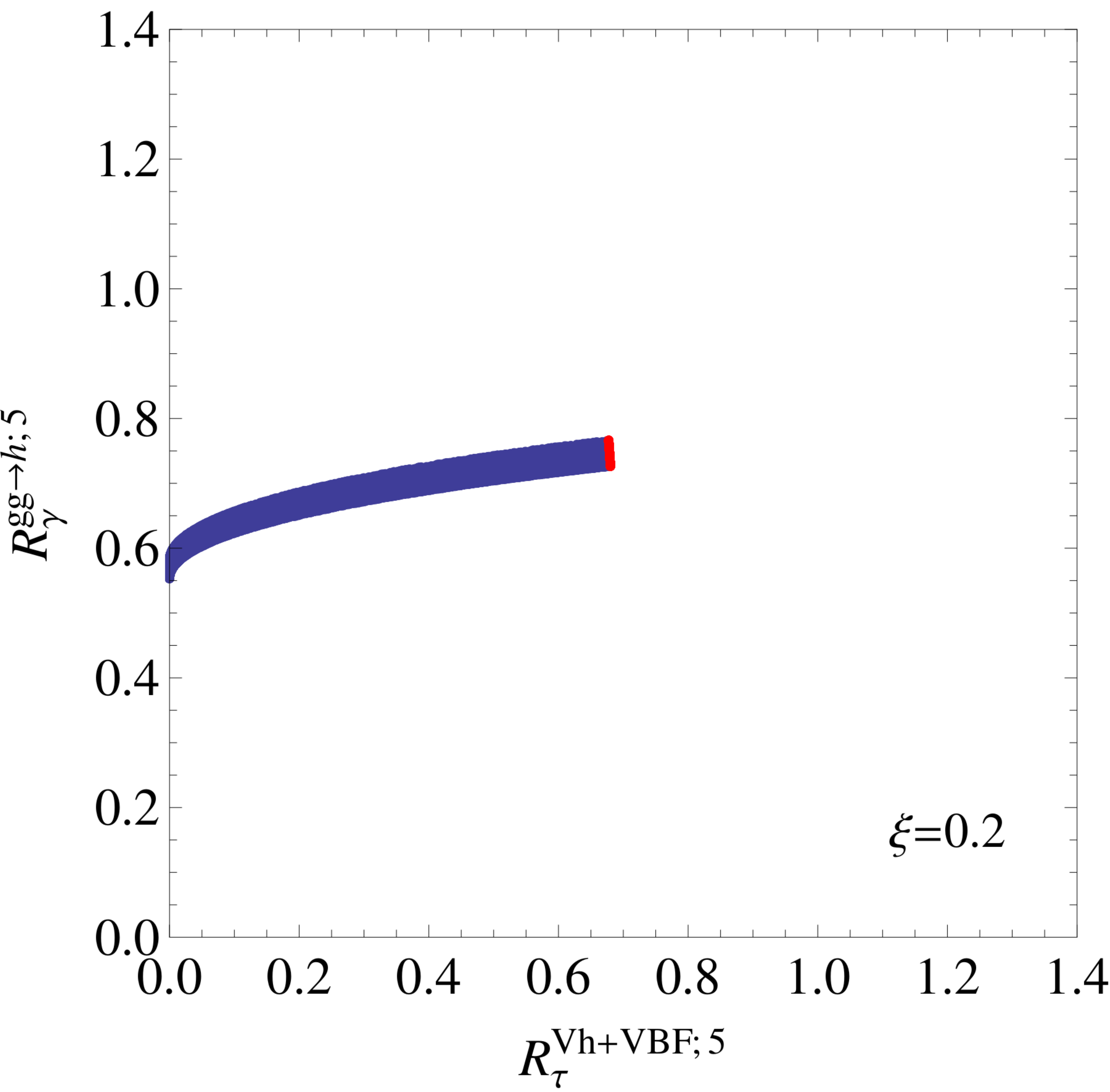}
\hspace{1.5cm}
\includegraphics[width=0.3\textwidth]{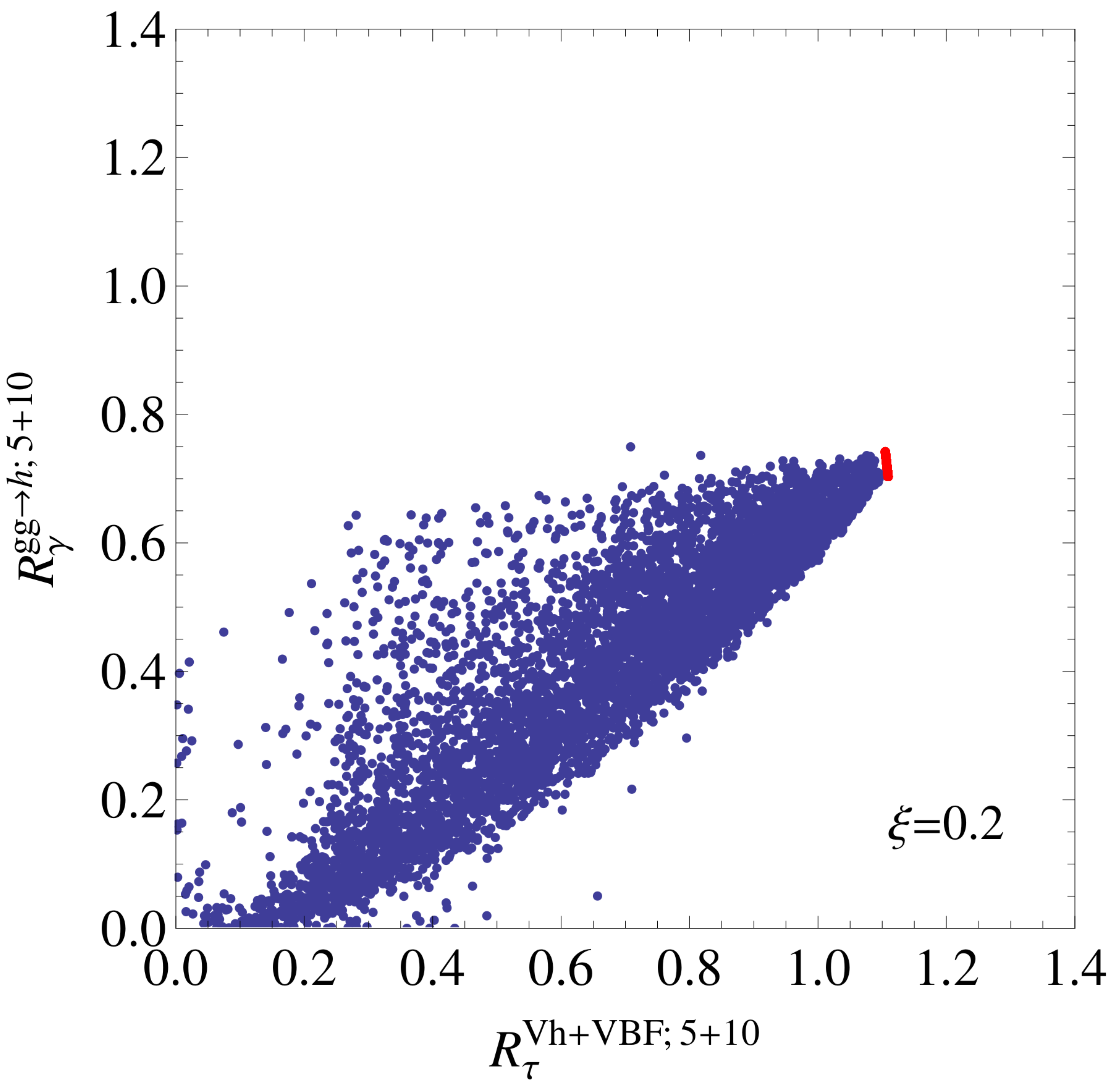}
\caption{\label{fig:gamtaun} Production cross section times decay for $gg\to h\to \gamma\gamma$  relative to the SM versus the same ratio for $h\to\tau\tau$ in VBF or $V h$ production, including the effects of the non-linearity of the Higgs as well as the additional correlations coming from the UV completion of the MCHM$_{5+10}$, see text and \cite{Carmona:2013cq} for more details.  }
\end{center}
\vspace{-0.5cm}
\end{figure}

Up to now, we have neglected the effects arising from the 
pseudo-Goldstone boson nature of the Higgs in the UV 
completions of our models. Considering this will lead to shifted Higgs couplings to 
the different fermions and gauge bosons of the spectrum. In the latter 
case, neglecting the (sub-leading) mixing of the SM gauge bosons to their composite 
counterparts, everything is fixed by the quantum numbers and the 
symmetry breaking defining the composite model \cite{Giudice:2007fh}
\begin{equation}
\kappa_W^{m}=\kappa_Z^{m}=\cos\left(\frac{v}{f}\right)\approx 
\sqrt{1-\xi},\qquad m=5,5+10,
\end{equation}
where we have defined $\xi=v^2/f^2$ as usual. The rescaling of fermions, on the other hand, depends on the particular representations. In the MCHM$_5$, we have the following universal shift of the Higgs coupling to SM fermions,
\begin{equation}
\kappa_f^5\to 
\kappa_f^5\cos\left(\frac{2v}{f}\right)/\cos\left(\frac{v}{f}\right)\approx 
\kappa_f^5(1-2\xi)/\sqrt{1-\xi}, 
\end{equation}
leading 
also to a suppression of the effective coupling to gluons,
\begin{eqnarray}
\kappa_{g}^m\approx\cos\left(\frac{2v}{f}\right)/\cos\left(\frac{v}{f}\right) \approx (1-2\xi)/\sqrt{1-\xi},\qquad m=5,5+10.
\end{eqnarray}
In the case of the MCHM$_{5+10}$ we get a different correction for the $\tau$ \cite{Azatov:2011qy}
\begin{eqnarray}
\label{eq:ktauc}
\kappa_{\tau}^{5+10}\to 
\kappa_{\tau}^{5+10}\cos\left(\frac{v}{f}\right)\approx 
\kappa_{\tau}^{5+10}\sqrt{1-\xi}
\end{eqnarray}
while all other fermionic couplings change analogously as in the MCHM$_5$.
%Finally, the change in $\kappa_\gamma^m$ can be worked out for both models by applying the replacements given above to 
%(\ref{eq:kappagamma}) (including the change in the Higgs coupling to $W^\pm$-bosons).
%
We have implemented all these additional corrections in our 
phenomenological study, employing $\xi=0.2$, to see to what extend the previous picture is 
changed. To this purpose we show in Figure \ref{fig:gamtaun} similar plots to the ones appearing in the left and middle panels of Figure \ref{fig:gamgamtautau}, taking into account the effects of the non-linearity of the Higgs (as well as the additional correlations coming from the UV completion, commented on above). Note that in the latter Figure we have also distinguished different production mechanisms for the processes considered; namely  $gg\to h$ for $h\to\gamma\gamma$ and $Vh+$VBF for $h\to\tau\tau$. We also show by red regions the predictions corresponding to zero mixing with the composite lepton sector. Comparing these with the effects due to the mixing with the vector-like leptons, we can see that the latter effect is still the dominant one for a composite $\tau$ scenario.

\section{Conclusions and Outlook}
\vspace{-1.8mm}

In this note, we have studied the impact of light custodians, present in composite Higgs models, on Higgs production and decay. In particular, we pointed out that the inclusion of a realistic 
lepton sector can lead to important in effects in the Higgs decays to two photons and $\tau$ leptons, whereas changes in the $tth$ coupling could be probed by comparing different production mechanisms.
Due to the effective (low energy) approach of our analyses,  it also provides interesting information about general viable vector-like fermion scenarios, independent of the UV completion.
Our focus on the third generation can then be motivated by its prominent role in the SM, suggesting a closer connection to electroweak symmetry breaking and new physics.
The predicted clear correlations allow for interesting tests of the setups.
We presented explicitly the impact of full composite completions, due to the non-linearity of the Higgs sector.

Finally, an obvious question is how composite the $\tau$ can be without generating a conflict with observables that are not protected by custodial symmetry. Tau compositeness  can for example lead to sizable effects in $(\bar e \gamma^\mu P_{L,R} e)\, (\bar \tau \gamma_\mu P_{R} \tau)$ transitions, due to the exchange of heavy composite resonances.
The coefficient of this operator is experimentally constrained to $C_{ee\tau\tau}\lesssim [(1-2)\,{\rm TeV}]^{-2}$ \cite{Raidal:2008jk}.
In the composite models considered here, it  is expected to scale as $C_{ee\tau\tau} \sim g^2 \times 1/\sqrt L \times \sqrt L \times (s_R^\tau)^2/m_\rho^2$, for $s_R^\tau \sim {\mathcal O}(1)$, where $L\approx35$ corresponds to an enhancement/suppression factor of the coupling, depending on the compositeness of the external fermions. Even for such values of $s_R^\tau$, the bound mentioned before is fulfilled for masses of the bosonic resonances as light as  $m_\rho \sim {\mathcal O}(1)$\,TeV.

Along similar lines, also the one-loop corrections to the $Z \tau \bar \tau$ vertex, due to triangles containing fermionic and  heavy bosonic resonances, are expected to be modest enough for reasonable regions of the parameter space, such as to still allow for interesting effects in Higgs physics without a conflict with precision tests. A corresponding detailed analysis will be presented elsewhere \cite{next}.

\vspace{-0.75mm}
\paragraph{Acknowledgements}

We acknowledge support from the SNF under contract 200021-143781.

\end{document}